\documentclass[10 pt,final,journal,letterpaper,oneside,twocolumn]{IEEEtran}
\usepackage{cite} 
\usepackage{algorithm}
\usepackage{amsmath,amssymb,amsfonts}
\usepackage{algorithmic}
\usepackage{graphicx}
\usepackage{subfigure}
\usepackage{textcomp}
\usepackage{amsthm}
\usepackage{lipsum}
\usepackage{mathtools}
\usepackage{cuted}
\usepackage{epstopdf}

\usepackage{caption,setspace}
\usepackage{xcolor}

\begin{document}

\title{\LARGE{Cooperative Backscatter NOMA with Imperfect SIC: Towards Energy \\Efficient Sum Rate Maximization in Sustainable 6G Networks}}
\author{Manzoor Ahmed, Zain Ali, Wali Ullah Khan, Omer Waqar, Muhammad Asif, Abd Ullah Khan,\\ Muhammad Awais Javed, and Fahd N. Al-Wesabi \thanks{Corresponding Author: Omer Waqar (owaqar@tru.ca).

Manzoor Ahmed is with the College of Computer Science and
Technology, Qingdao University, Qingdao 266071, China (email: manzoor.achakzai@gmail.com).

Zain Ali is with Department of Electrical and Computer Engineering, University of California, Santa Cruz, USA (email: zainalihanan1@gmail.com). 

Wali Ullah Khan is with Interdisciplinary Center for Security, Reliability and Trust (SnT), University of Luxembourg, 1855 Luxembourg City, Luxembourg (email: waliullah.khan@uni.lu). 

Omer Waqar is with the Department of Engineering, Thompson Rivers University (TRU), British Columbia, Canada (email: owaqar@tru.ca).

Muhammad Asif is with the with the
Guangdong Key Laboratory of Intelligent Information Processing, College
of Electronics and Information Engineering, Shenzhen University, Shenzhen,
Guangdong 518060, China. (email: masif@szu.edu.cn).

Abd Ullah Khan is with the Department of Computer Science,
National University of Science and Technology Balochistan Campus, Quettta (email: abdullah@nbc.nust.edu.pk).

Muhammad Awais Javed is with the Department of Electrical and Computer Engineering, COMSATS University Islamabad, Islamabad 45550, Pakistan (email:
awais.javed@comsats.edu.pk).

Fahd N. Al-Wesabi is with the Department of Computer Science, College of Science \& Art at Mahayil, King Khalid University, Saudi Arabia
\& Faculty of Computer and IT, Sana’a University, Yemen (email: falwesabi@kku.edu.sa)
}}%

\markboth{}%
{Shell \MakeLowercase{\textit{et al.}}: Bare Demo of IEEEtran.cls for IEEE Journals}

\maketitle

\begin{abstract}
The combination of backscatter communication with non-orthogonal multiple access (NOMA) has the potential to support low-powered massive connections in upcoming sixth-generation (6G) wireless networks. More specifically, backscatter communication can harvest and use the existing RF signals in the atmosphere for communication, while NOMA provides communication to multiple wireless devices over the same frequency and time resources. This paper has proposed a new resource management framework for backscatter-aided cooperative NOMA communication in upcoming 6G networks. In particular, the proposed work has simultaneously optimized the base station's transmit power, relaying node, the reflection coefficient of the backscatter tag, and time allocation under imperfect successive interference cancellation to maximize the sum rate of the system. To obtain an efficient solution for the resource management framework, we have proposed a combination of the bisection method and dual theory, where the sub-gradient method is adopted to optimize the Lagrangian multipliers. Numerical results have shown that the proposed solution provides excellent performance. When the performance of the proposed technique is compared to a brute-forcing search technique that guarantees optimal solution however, is very time-consuming, it was seen that the gap in performance is actually 0\%. Hence, the proposed framework has provided performance equal to a cumbersome brute-force search technique while offering much less complexity. The works in the literature on cooperative NOMA considered equal time distribution for cooperation and direct communication. Our results showed that optimizing the time-division can increase the performance by more than 110\% for high transmission powers. 
\end{abstract}


\IEEEpeerreviewmaketitle

\section{Introduction}

The upcoming sixth-generation (6G) systems are expected to connect billions of communication devices all over the world \cite{giordani2020toward,9521550}. Most promising 6G technologies are artificial intelligence/machine learning \cite{shome2021federated,ali2021artificial}, reconfigurable intelligent surfaces \cite{basar2020reconfigurable}, backscatter communication \cite{van2018ambient}, non-orthogonal multiple access (NOMA) \cite{khan2020efficient}, blockchain \cite{sekaran2020survival,jameel2020reinforcement11}, Tera-hertz communication \cite{oleiwi2022cooperative}, and simultaneous wireless information and power transfer \cite{oleiwi2022cooperative11}. These technologies will integrate to the current communication networks such as unmanned aerial vehicles \cite{haider2021energy}, intelligent transportation systems \cite{jameel2020efficient}, cognitive radio networks \cite{tanveer2021enhanced}, Internet of Things \cite{ali2021efficient}, device to device communication \cite{yu2021optimal}, and physical layer security \cite{jameel2019secrecy}. However, the main challenges would be the spectrum scarcity and limited energy reservoirs, specially for those systems using conventional orthogonal multiple access (OMA) protocol \cite{khan2019efficient}. In this regard, researchers in academia and industry are studying the above new technologies. 

\begin{figure}[!t]
\centering
\includegraphics [width=0.45\textwidth]{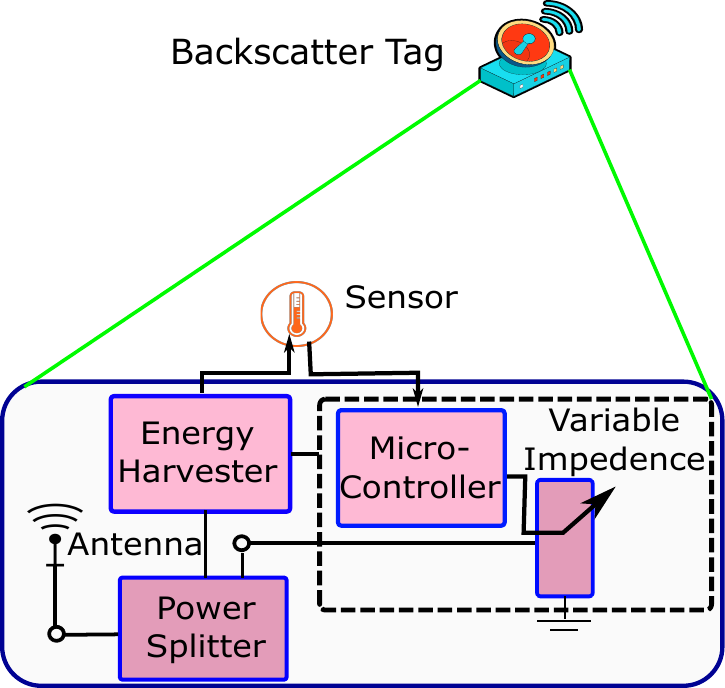}
\caption{Basic architecture of backscatter sensor device}
\label{f1}
\end{figure}

Backscatter communication and NOMA are two examples of emerging technologies that enhance spectrum and energy efficiency of 6G systems \cite{jameelMag}. Further, NOMA has been shown to outperform OMA protocol \cite{8861078}. With the help of the ambient energy harvesting approach, backscatter communication allows sensor devices to transmit data towards surrounding users by reflecting and modulating radio frequency (RF) signal \cite{li2021physical}. The basic architecture of backscatter sensor device can be seen in Fig. \ref{f1}. One the other hand, NOMA enables the transmission of multiple users over the same spectrum/time resources using the superposition coding, and successive interference cancellation (SIC) techniques \cite{liu2017non,9479745}. The performance of backscatter communication has been previously studied in OMA networks \cite{jameel2021multi,khan2021learning,jameel2019towards}. The integration of NOMA with backscatter communication is a hot topic and some works in literature have investigated different problems related to backscatter communication in NOMA wireless networks. 

\subsection{Recent Advances in NOMA Backscatter Communication Networks}
Cooperative communication has been shown to improve the performance of communication systems significantly \cite{8437135}. In cooperative communication, either a dedicated device (called relay) is used to forward the data of a specific user \cite{9314919} or a communicating user cooperates by relaying the data to other users\cite{ali2021fair}. The works in \cite{9314919} and \cite{ali2021fair} optimized power allocation in cooperative NOMA systems to maximize the sum rate of the system and to achieve fairness, respectively. For sum rate maximization in cooperative NOMA based device-to-device communication, Jiang {\em et al.} \cite{8641249} proposed an optimization framework. Kim {\em et al.} \cite{8038077}  proposed a power optimization algorithm to achieve the maximum capacity scaling in a cooperative NOMA scenario. Further, Reference \cite{khan2019maximizing} explored an optimization problem to enhance the secrecy rate of NOMA cooperative communication. Of late, the work of authors in \cite{oleiwi2022cooperative} have proposed a cooperative simultaneous wireless information and power transfer in NOMA-enabled terahertz communications to improve energy and spectral efficiency of the system.

Recently, researchers have studied the integration of NOMA with backscatter communication in next generation wireless networks \cite{8439079}. For instance, Zhang {\em et al.} \cite{8636518} provided the closed-form expressions for the outage probability and ergodic capacity in backscatter-aided NOMA symbiotic system. Khan {\em et al.} \cite{9345447} proposed backscatter-aided vehicular-to-everything network and jointly optimized the transmit power of base station (BS) and roadside units to maximize the sum capacity of the NOMA system. The work in \cite{8851217} jointly optimized the time allocation and reflection coefficient of the backscatter tag to maximize the minimum throughput of the NOMA Internet of things network. To maximize the energy efficiency, Xu {\em et al.} \cite{9223730} explored a joint optimization framework of transmit power and reflection coefficient in backscatter-aided NOMA network. The authors of \cite{9383801} derived a closed-form expression for bit error rate in backscatter-aided NOMA network under imperfect SIC. Besides, the authors of \cite{9328505} investigated the optimization problem of transmit power and reflection coefficient to maximize the sum rate of backscatter-aided NOMA network under imperfect SIC. Reference \cite{9319204} calculated the closed-form expressions of intercept and outage probability for backscatter-aided NOMA system under residual hardware impairments and imperfect channel state information (CSI) and SIC. To improve the energy efficiency of the system, the work in \cite{khan2021energy} investigated a new optimization approach for efficient power allocation and reflection coefficient under imperfect SIC. Asim {\em et al.} \cite{ihsan2021energy} proposed an uplink optimization framework for NOMA backscatter sensor communication under channel estimation errors in intelligent transportation systems. The research work in \cite{khan2021joint} has maximized the spectral efficiency of NOMA backscatter communication networks. Manzoor {\em et al.} \cite{ahmed2021backscatter} also maximized energy efficiency of multi-cell NOMA backscatter sensor networks under imperfect SIC. Of late, the authors of \cite{khan2021integration} also considered imperfect SIC in multi-cell NOMA backscatter communication to maximize the spectral efficiency of the network.

\subsection{Motivation and Contributions}
Most of the above literature \cite{8636518,9345447,8851217,9223730,ihsan2021energy,khan2021joint,9314919,
ali2021fair,8641249,8038077} assumes perfect SIC in their systems which is impractical. The works in \cite{9383801,9328505,9319204,khan2021energy,ahmed2021backscatter,khan2021integration} consider imperfect SIC, however, cooperation among the communicating users was not considered. Further, the authors in \cite{9314919,ali2021fair,8641249,8038077,khan2019maximizing} just optimized the power allocation while considering equal time allocation on both hops. To the best of our knowledge, the problem of resource management that simultaneously optimizes the transmit power of BS and relaying node, the reflection coefficient of backscatter tag, and time allocation in cooperative NOMA network under imperfect SIC has not yet been investigated. To fill this bridge, we aim to provide a resource management framework to maximize the sum rate of backscatter-aided cooperative NOMA network under imperfect SIC. Closed-form solutions are derived by dual theory and KKT conditions, where numerical results demonstrate the superiority of joint optimization with backscattering enabled system over the conventional fix time cooperation and communication without any backscattering. The main contributions of our work can also be summarized as:
\begin{enumerate}
\item This paper considers a new optimization framework for a backscatter-aided NOMA cooperative communication, where a BS transmits superimposed data to two NOMA users. This work also considers that the near user performs cooperation by relaying data to a far user. Meanwhile, a backscatter tag also receives superimposed signal from BS and cooperative user. The backscatter tag modulates its information and then reflect it towards both users. Thus, users also act as readers. The communication process takes two-time slots. In the first time slot, BS transmits to both users, and the backscatter tag reflects the received signal of BS towards both users. In the second time slot, near user cooperate by relaying data to a far user, and the backscatter tag reflects the relaying signal to both users by adding useful information. 
\item  We formulate a new optimization problem to maximize the sum rate of backscatter-aided NOMA cooperative communication under imperfect SIC decoding while satisfying various practical constraints. In particular, we jointly optimize the transmit power of BS, cooperative power of near user, the reflection coefficient of backscatter tag, and time allocation of both time slots while ensuring the minimum data rate of both users. The formulated problem is non convex optimization, and joint optimization cannot be designed to obtain the solution. Thus, we adopt the bisection method and dual theory to obtain an efficient solution, where the values of dual variables are iteratively updated.
\item To see the benefits of backscatter communication, the proposed work also provides the optimization of conventional cooperative NOMA communication without backscattering and the Brute Force Search technique for comparison. Numerical results are plotted using Monte Carlo simulations. Results demonstrate that the proposed optimization approach obtains a higher sum rate than the other benchmark scheme and converges within a reasonable number of iterations.
\end{enumerate}

The remaining of our work can be organized as follows. Section II will provide the system model, various assumptions, and optimization problem. Section III will discuss different steps of proposed optimization solution to enhance the sum rate of the system. Section IV will present and discuss the numerical results based on Monte Carlo simulations while Section V will conclude this paper with some interesting research directions. Description of important parameters is provided in Table I.

\section{ System Model and Problem Formulation}

\begin{table}[]
\begin{tabular}{|l|l|}
\hline 
\!\!Parameter\!\!     & Description                                                                                   \\ \hline \hline 
$P$           & Available power at the BS                                                                     \\ 
$P_r$          & Available power at $U_1$ for relaying                                                         \\ 
$\Lambda$     & \begin{tabular}[c]{@{}l@{}}Faction of $P$ allocated for transmission\\  of $U_1$\end{tabular} \\ 
($1-\Lambda$) & \begin{tabular}[c]{@{}l@{}}Faction of $P$ allocated for transmission\\  of $U_1$\end{tabular} \\ 
$T$           & Time allocated for direct transmission                                                        \\ 
($1-T$)       & Time allocated  for cooperation                                                               \\ 
$\phi_1$      & Reflection coefficient in first time slot ($T$)                                                 \\ 
$\phi_2$      & Reflection coefficient in second time slot (1-$T$)                                              \\ 
$g_1$         & Channel gain from BS to $U_1$                                                                 \\ 
$g_2$         & Channel gain from BS to $U_2$                                                                 \\ 
$g_3$          & Channel gain from BS to Backscatter Tag                                                       \\ 
$h_1$          & Channel gain from $U_1$ to $U_2$                                                              \\ 
$h_2$         & Channel gain from $U_1$ to Backscatter Tag    \\ 
$f_1$         & Channel from Backscatter Tag to $U_1$                 \\ 
$f_2$         & Channel from Backscatter Tag to $U_2$                      \\ 
$\sigma^2$    & Variance of additive white Gaussian noise                                  \\ \hline
\end{tabular}
\caption{Description of important parameters}
\end{table}
\begin{figure}[!t]
\centering
\includegraphics [width=0.45\textwidth]{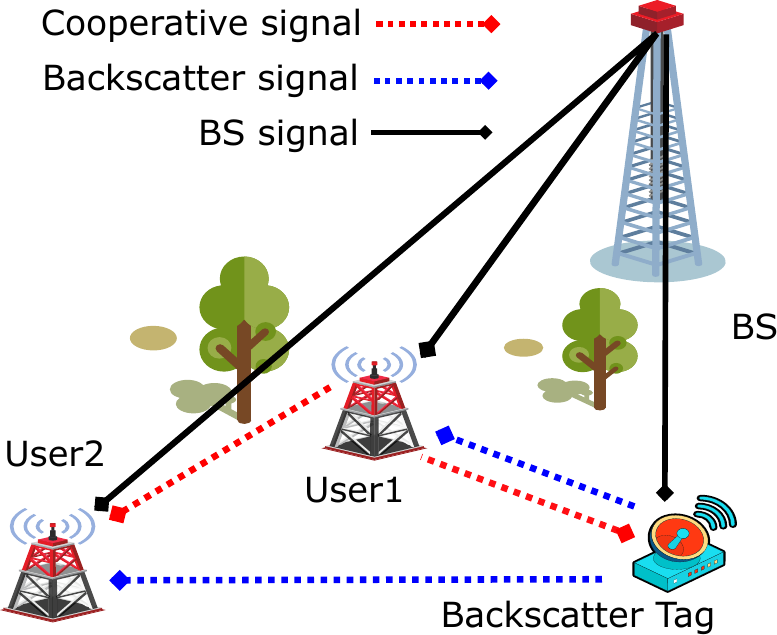}
\caption{System model of backscatter-aided cooperative NOMA communications.}
\label{f11}
\end{figure}
We consider a  backscatter-aided downlink cooperative NOMA communication as shown in Fig. \ref{f11}, where a BS serves data to two users (end users) in the cell\footnote{Although the work considers a single cell system, the proposed solution can be implemented directly in multi-cell networks. For example, the same spectrum resources can be reused in a multi-cell scenario such that each cell will cause inter-cell interference to other neighboring cells. In such a scenario, optimization of transmit power, user association, and time allocation become more important due to inter-cell interference. This is an exciting research topic, and we plan to investigate it in our future studies.}. The coverage area also contains a backscatter tag that sends data to both the receivers using the existing radio signals. More specifically, the backscatter tag also receives the superimposed signal of BS and harvests some energy from the signal to operate the circuit. Then modulate useful data and reflect it towards users using the reflection coefficient. This work considers that all the nodes are equipped with a single antenna and that the perfect channel state information is available at the BS. The channels between different links are independent and identical and undergo Rayleigh fading. The user closer to the BS has better channel conditions is represented as $U_1$, and the far user with a comparatively much lower value of channel gain is denoted as $U_2$. We have assumed that $U_1$ is also closer to the backscatter tag than $U_2$. The available transmission time is divided into two slots. In the first time slot ($T$), the BS transmits data to both users; the signal from the BS is also received and reflected by the backscatter tag. The $U_1$ employs SIC to decode its own signal; during the process of SIC, the $U_1$ first decodes the signal of $U_2$ and then subtracts this decoded data from the received signal to decode its own signal. In the remaining time (1-$T$), the $U_1$ cooperates by relaying the decoded data to $U_2$. This relayed data is also reflected by the backscatter tag. Based on the above discussion and consideration, the signal received at $U_1$ and $U_2$ in the first time slot is given as:
\begin{align}
y_{U1,1}=&\sqrt{g_1}\Big(\sqrt{P \Lambda} x_{U1}+\sqrt{P (1-\Lambda)} x_{U2}\Big)+\sqrt{\phi_1 g_3 f_1}\nonumber\\&\Big(\sqrt{P \Lambda} x_{U1}+\sqrt{P (1-\Lambda)} x_{U2}\Big)z(t),
\end{align}
\begin{align}
y_{U2,1}=&\sqrt{g_2}\Big(\sqrt{P \Lambda} x_{U1}+\sqrt{P (1-\Lambda)} x_{U2}\Big)+\sqrt{\phi_1 g_3 f_2}\nonumber\\&\Big(\sqrt{P \Lambda} x_{U1}+\sqrt{P (1-\Lambda)} x_{U2}\Big)z(t),
\end{align}
where $z(t)$ is the signal added by the backscattering tag, with $\mathbb{E}[|z(t)|^2]=1$ and $P$ denotes the total available power at the BS. $x_{U1}$ is the data symbol of $U_1$ and $x_{U2}$ represents the data symbol of $U_2$, respectively. $\Lambda$ and ($1-\Lambda$) are the fractions of $P$ allocated for the signal of $U_1$ and $U_2$ by the BS. $g_1$, $g_2$ , $g_3$, $h_1$ and $h_2$ are the channel gains from BS to the $U_1$, $U_2$, from BS to backscatter tag, from the $U_1$ to $U_2$ and from $U_1$ to the backscatter tag, respectively. Similarly, $f_1$ and $f_2$ are the channel gains from backscatter tag to $U_1$ and $U_2$. $\phi_1$ is the reflection coefficients in the first time slot. Then, in the second time slot, the signal received by $U_2$ is given as:
\begin{align}
y_{U2,1}=&\sqrt{P_r h_1} x_{U2}+\sqrt{\phi_2 P_r h_2 f_2} x_{U2}z(t),
\end{align}
where $P_r$ is the power invested by $U_1$ for relaying data to $U_1$.
Note that, we have considered that on receiving the signal from the backscattering tag, $U_1$ successfully perform SIC to decode its symbol without any interference. However, as the BS is much far compared to the backscattering tag, hence the SIC is imperfect and the $U_1$ faces some interference while decoding this symbol. Then, the achievable rates at the $U_1$ and $U_2$ in the first time slot are given as $T R_{1}$ and $T R_{2}$. The rate due to relaying at the $U_2$ is $(1-T) R_{3}$, and the rate of decoding of the $U_2$ symbol at the $U_1$ in the first time slot is $T \overline{R_1}$. The values of $R_1$, $R_2$, $R_3$ and $\overline{R_1}$ are computed as:
\begin{align}
R_{1}= \log_{2}\Bigg(1+\dfrac{P \Lambda^*(g_{1}+\phi_{1}^*f_{1}g_{3})}{P(1-\Lambda^*)g_{1}\beta+\sigma^2}\Bigg),
\end{align}
\begin{align}
R_{2}= \log_{2}\Bigg(1+\dfrac{P (1-\Lambda^*)(g_{2}+\phi_{1}^*f_{2}g_{3})}{P\Lambda^* (g_{2}+\phi_{1}^* f_{2} g_{3})+\sigma^2}\Bigg),
\end{align}
\begin{align}
R_{3}=\log_{2}\Bigg(1+\dfrac{P_r^*(h_{1}+\phi_{2}^* f_{2} h_{2})}{\sigma^2}\Bigg),
\end{align}
\begin{align}
\overline{R_{1}}=\log_{2}\Bigg(1+\dfrac{P (1-\Lambda^*)(g_{1}+\phi_{1}^*f_{1}g_{3})}{P\Lambda^*(g_{1}+\phi_{1}^*f_{1}g_{3})+\sigma^2}\Bigg),
\end{align}
where $\sigma^2$ is the variance of additive white Gaussian noise. As this work considers the practical scenario of imperfect SIC, $\beta$ signifies the fraction of interference faced by the $U_1$ while decoding its own data. Then, the problem of maximizing the sum rate is written as:  
\begin{alignat}{2}
\mathcal P: \ & \underset{{(T,\Lambda,\phi_{1},\phi_{2},P_r)}}{\text{max}}\ T R_1+T R_2+ (1-T) R_3\label{8}\\
s.t.&
\begin{cases}
 \mathcal C_1: T R_1\geq R_{min}, \nonumber\\
 \mathcal C_2: T R_2+(1-T)R_3\geq R_{min}, \nonumber\\
 \mathcal C_3: T \overline{R_1}\geq T R_2+(1-T)R_3, \nonumber\\
 \mathcal C_4: P_r\leq Pr_{max}, \forall i,j, \nonumber\\
 \mathcal C_5: 0\leq T\leq 1,0\leq \phi_{1}\leq 1,0\leq \phi_{2}\leq 1,0\leq \Lambda\leq 1, \nonumber
\end{cases}
\end{alignat}
where the objective in (8) is to maximize the sum rate of the system. The first two constraints $\mathcal C_1$ and $\mathcal C_2$ ensure that the minimum rate requirement of the users is satisfied, where $R_{min}$ denotes the rate required by the users. Similar to \cite{zhao2010power}, the third constraint $\mathcal C_3$ guarantees that the cooperation required is fulfilled. Then, $\mathcal C_4$ makes sure that power allocation at $U_1$ will follow the power budget, where $Pr_{max}$ is the battery capacity of the user. Finally, constraint $\mathcal C_5$ ensures that the values of time and reflection coefficients will remain within the practical range. 
\section{Proposed Solution}

The considered problem $\mathcal P$ is a multi-variable complex optimization problem as the objective function is not concave in all variables. Thus, a joint optimization framework cannot be designed to obtain the solution. For optimization, we take into consideration the independent impact of all parameters on the objective function and propose solutions subject to the nature of the objective function with respect to the specific parameters. 

First we investigate the efficient value of reflection coefficient at backscatter tag for both time slots. As the objective function is concave with respect to $\phi_1$, $\phi_2$ and $P_r$. We employ duality theory to find the solution for these variables, where the Lagrangian of the problem is written as:
\begin{align}
&L=T R_1+T R_2+(1-T) R_3+\lambda_{1}\Big(T R_1- R_{min}\Big)+\lambda_{2}\nonumber\\&\Big( T R_2+(1-T)R_3- R_{min}\Big)+\mu\Big(T \overline{R1}- T R_2-(1-T)\nonumber\\&R_3\Big)+\eta(Pr_{max}-P_r)+\zeta_{1}(1-\phi_{1})+\zeta_{2}(1-\phi_{2}),
\end{align}
where $\lambda_1,\lambda_2, \mu,\eta,\zeta_1$ and $\zeta_2$ are the Lagrangian multipliers. Then applying Karush–Kuhn–Tucker (KKT) conditions and differentiating with respect to $\phi$ results in:
\begin{align}
\phi_{1}^5\theta_{5}+\phi_{1}^4\theta_{4}+\phi_{1}^3\theta_{3}+\phi_{1}^2\theta_{2}+\phi_{1}\theta_{1}+\theta=0, \label{o8}
\end{align}
\noindent the values of $\theta,\theta_{1},\theta_{2}\text{ and }\theta_{3}$ are given in (\ref{n13}), (\ref{n14}), (\ref{n15}), (\ref{n16}) and (\ref{n17}). Please refer to Appendix A.
Note that, (\ref{o8}) is a monic polynomial of degree 5, the solution can be easily found by using conventional methods or by employing the solvers provided in MATLAB, Mathematica etc. 

Similarly, employing KKT conditions and differentiating with respect to $\phi_2$, the solution IS given as:
\begin{align}
\phi_{2}^*= \max(0,\omega),\label{o15}
\end{align}
where
\begin{align}
\omega=\dfrac{-f_{2} g_{3} (1 + \lambda_{2} - \mu) P_r (-1 + T) - (h_{1} P_r + \sigma^2) \zeta_{2}}{f_{2} g_{3} P_r \zeta_{2}},
\end{align}

Next we compute the efficient value of relayed power at user. For the solution of $P_r$, we take advantage of the fact that the first two terms in $\mathcal C_3$ of problem (8) are constant with respect to $P_r$. Thus, we write the constraint in $\mathcal C_3$ of problem (8) as
\begin{align}
\mathcal{C'}_3= (2^\psi -1)\sigma^2\geq P_r (h_1+\phi_2 f_2 h_2),
\end{align}
where $\psi= (T \overline{R_1}- T R_2)/(1-T)$. After this transformation, next applying KKT conditions and differentiating the Lagrangian with respect to $P_r$, gives us:
\begin{align}
&P_r^*=\max(0,\Psi)
\label{o116}
\end{align}
\begin{align}
&\Psi=\nonumber\\&\dfrac{ h_{1} (1 + \lambda_{2}) (1 -T)+  f_{2} g_{3} (1 + \lambda_{2}) \phi_{2} (1 - T)-(\eta + \mu) \sigma^2   }{(\eta + \mu) (h_{1} + f_{2} g_{3} \phi_{2})}, \label{o16}
\end{align}

\begin{algorithm}[!t]
\caption{Solution for $T^*$ using Bisection method}
\begin{enumerate}
\item \textbf{Initialize:} system parameters and variables.
\item \textbf{Calculate} $\phi_1^*,\phi_2^*,\Lambda^*$ and $P_r^*$ for $T=0.5-\Delta$
\item \textbf{Set} $\tau_L=0$ and $\tau_U=1$, Rbest=$\omega(\phi_1^*,\phi_2^*,\Lambda^*,P_r^*,T)$
\item \textbf{while} $|\tau_L-\tau_U|>$ $\epsilon$
\item \textbf{Set} $\tau=\dfrac{\tau_L+\tau_U}{2}$
\item \textbf{Calculate} $\phi_1^*,\phi_2^*,\Lambda^*,P_r^*$ for $T=\tau$
\item \textbf{if}  $\omega(\phi_1^*,\phi_2^*,\Lambda^*,P_r^*,\tau)>$Rbest
\item \textbf{set} Rbest=$\omega(\phi_1^*,\phi_2^*,\Lambda^*,P_r^*,\tau)$, $\tau_L=\tau$, $T^*=\tau$
\item \textbf{else}
\item \textbf{Set} $\tau_U=\tau$
\item \textbf{end while}
\item \textbf{Return} $T^*$
\end{enumerate}
\label{algo1}
\end{algorithm}
Now, we calculate the power allocation coefficient at BS. For perfect SIC, i.e, $\beta=0$, the objective function in problem (8) is concave with respect to $\Lambda$. Thus the solution can be found using techniques proposed in Reference \cite{9261140}. However, for the considered imperfect SIC case, the eigenvalue value of the Hessian of the objective with respect to $\Lambda$ is given as $EV=\dfrac{T \kappa}{\chi}$, where the values of $\kappa$ and $\chi$ are:\footnote{For this purpose, first we need to find the Hessian of the function, then the matrix is transformed into an upper triangular matrix. The diagonal entries of the upper triangular matrix are the eigenvalues of the function. However, for the considered case of single variable $\Lambda$, the double derivative of the function gives us the eigenvalues and provides the information about the convexity of the function.}
\begin{align}
&\kappa=a^2 (c^4 (-1 + \Lambda)^4 + \sigma^6 (2 b L + \sigma^2) + 2 c \sigma^4 (b (3 - 2 \Lambda) \Lambda -\nonumber\\& 2 (-1 + \Lambda) \sigma^2) -  2 c^3 (-1 + \Lambda) (b \Lambda (1 - \Lambda + \Lambda^2) + 2 (-1 +\nonumber\\& \Lambda)^2 \sigma^2) +  c^2 (b^2 \Lambda^4 + 2 b \Lambda (3 - 4 \Lambda + 2 \Lambda^2) \sigma^2 + 6 (-1 + L)^2 \nonumber\\&\sigma^4)) -2 a b \Lambda \sigma^2 (c + \sigma^2) (2 c (c (-1 + \Lambda) - \sigma^2) + b (c - 2 c \Lambda\nonumber\\& + s)) -  b \sigma^4 (c + \sigma^2) (2 c (c (-1 + \Lambda) - \sigma^2) + b (c - 2 c \Lambda + \nonumber\\&\sigma^2)),
\end{align}
\begin{align}
& \chi=(a \Lambda + \sigma^2)^2 (c - c \Lambda + \sigma^2)^2 (c + b \Lambda - c \Lambda + \sigma^2)^2
\end{align}
where $a=P(g_{2}+\phi_{1}f_{2}g_{3}),$ $b=P(g_{1}+\phi_{1}f_{1}g_{3})$ and $c=Pg_{1}\beta$. The eigenvalue is positive for $0\leq\Lambda<1$ and $ a>>\sigma^2$. Hence, the objective of the problem is convex with respect to $\Lambda$. For $\omega(\rho)$, denoting the value of the objective at $\Lambda=\rho$, convexity implies that we have $\omega(u \rho_1+(1-u) \rho_2)\leq u \omega(\rho_1)+(1-u)\omega( \rho_2)$ for $u\in [0,1]$. This shows that the maximum value of the objective function lies on either of the two extremes of $\Lambda$.\footnote{This means that the maximum value of the sum rate is achieved either when the $\Lambda$ is assigned the maximum value while satisfying all the constrains or when the value of $\Lambda$ is set to the minimum, such that, all the constraints are satisfied.} Since $U_1$ is closer to the BS compared to $U_2$, thus, the channel gains can be sorted as $g_1>g_2$. This shows that $\mathcal C_3$ in problem (8) is always satisfied for any value of $\Lambda$, and the feasibility of the constraint depends only on the value of $P_r$. Thus the value of $\Lambda$ is bounded by the rate requirements of both users. The lower bound due to the rate requirement of $U_1$ is given by: 
\begin{align}
&\alpha_{L1}=\dfrac{( 2^{R_{min}/T}-1) (\beta g_{1} P + \sigma^2)}{P (g_{1} - \beta g_{1} + \beta 2^{R_{min}/T} g_{1} + f_{1} g_{3} \phi_{1})},\label{o19}
\end{align}
similarly, for given $P_r^*$ the upper bound is written as:
\begin{align}
    &\alpha_{U1}=\dfrac{P(g_{2}+f_{2}g_{3}\phi_{1})-(2^{(R_{min}-(1-T)Rr)/T}-1)\sigma^2}{(2^{(R_{min}-(1-T)Rr)/T}) P(g_{2}+f_{2}g_{3}\phi_{1})},\label{o21}
\end{align}
where $Rr=(1-T)\log_{2}\Bigg(1+\dfrac{P_r^*(h_{1}+\phi_{2}^* f_{2} h_{2})}{\sigma^2}\Bigg)$. Then, the lower bound and upper bounds are given as $\alpha_{L}=\min(\alpha_{L1},1)$ and $\alpha_{U}=\max(\alpha_{U1},0)$, respectively. After this, calculating $\Lambda^*$ is straightforward, if $\omega(\alpha_{L})>\omega(\alpha_{U})$, then $\Lambda^*=\alpha_{L}$, otherwise, $\Lambda^*=\alpha_{U}$. Here, $\omega(\rho)$ represents the value of objective function at $\Lambda=\rho$. For finding the solution of Lagrangian multiplier, we employ subgradient method where in each iteration, the values of 
the dual variables are updated as:

\begin{align}
&\lambda_{1}^{t+1}=\lambda_{1}^{t}-\delta\Big(T R_1- R_{min}\Big),
\end{align}
\begin{align}
&\lambda_{2}^{t+1}=\lambda_{2}^{t}-\delta\Big( T R_2+(1-T)R_3- R_{min}\Big),
\end{align}
\begin{align}
&\mu^{t+1}=\mu^{t}-\delta\Big(T \overline{R_1}- T R_2-(1-T)R_3\Big),
\end{align}
\begin{align}
&\eta^{t+1}=\eta^{t}-\delta(Pr_{max}-P_r),
\end{align}
\begin{align}
&\zeta_{1}^{t+1}=\zeta_{1}^{t}-\delta(1-\phi_{1}),
\end{align}
\begin{align}
&\zeta_{2}^{t+1}=\zeta_{2}^{t}-\delta(1-\phi_{2}),
\end{align}
where $t$ is the iteration index and $\delta$ represents the step size. 

Finally, we optimize $T$. Note that the problem of optimizing $T$ is a linear programming problem. For an objective function concave in $\Lambda$, this linear problem can be solved easily by using the simplex method. However, as the considered problem is convex with respect to $\Lambda$, this shows that at $\Lambda^*$, the rate of a user will be tightly bound\footnote{The convex objective function indicate that at optimal value of $\Lambda$, the rate of either $U_1$ or $U_2$ will be equal to $R_{min}$.}. Thus, for the given $\Lambda^*$, optimizing $T$ would have no impact. In this scenario, we are required to optimize $T$ and then calculate the values of $\phi_1^*,\phi_2^*,\Lambda^*$ and $P_r^*$ for the given $T^*$. Hence, we employ bisection method to optimize the value of $T$. The detailed steps involved in this optimization method are given in Algorithm \ref{algo1}. 

In the bisection method, first all the system parameters are initialized. In the second step, for the given T, where $T=0.5-\Delta$ (for $\Delta$ be a small positive number close to zero), the values of $\phi_1^*,\phi_2^*,\Lambda^*$ and $P_r^*$ are calculated. In third step, the bounds of the bisection method are initialized, where the lower bound ($\tau_L$) and the upper bound are set equal to 0 and 1, respectively. Then, we calculate the value of objective function for the given values of $\phi_1^*,\phi_2^*,\Lambda^*,P_r^*$ and $T$. The function $\omega(\phi_1^*,\phi_2^*,\Lambda^*,P_r^*,T)$ signifies the value of objective function for the given parameters. We set Rbest equal to this rate, where Rbest denotes the maximum value of sum rate achievable till now. In step 4, if the difference between $\tau_L$ and $\tau_U$ is greater than the permitted error $\epsilon$, the expected solution of $T$ represented as $\tau$ in step 5 is calculated. Next in step 6, the values of $\phi_1^*,\phi_2^*,\Lambda^*$ and $P_r^*$ for $T=\tau$ are calculated. If the value of $\omega(.)$ is greater than Rbest for these parameters, then we update Rbest, $T^*$ and set the lower bound $T_L$ equal to $\tau$. Otherwise, the upper bound is set equal to $\tau$ in step 10. Steps 5 to 10 are repeated until the difference between $\tau_L$ and $\tau_U$ falls below the permitted error.

The computational complexity of the proposed scheme can be given as $O(BIC)$, where $I$ denotes the number of iterations required by the Duality based method to provide the solution, $B$ represents the steps taken by the bisection method to reach the best solution of time allocation, and $C$ is the computational complexity of computing equations (11), (14), (18)-(25). Note that in the case where time allocation is not optimized, the computational complexity of the framework will be $O(IC)$. In the case of multi-cell system, the complexity will remain unchanged, because the optimization framework would run in each cell independently of all the other cells in the network.
\begin{table}[!t]
\caption{Simulation parameters and values}
\begin{tabular}{|c|c|} 
\hline 
Parameter & Value  \\
\hline\hline
Power budget of BS, $P$ & 40 dBm \\\hline
Reflection coefficient of backscatter  & $0\leq\xi\leq 1$ \\\hline
Channel type & i.i.d Raleigh fading  \\\hline
Cooperation power, $Pr_{max}$ & 30 dBm \\\hline
Imperfect SIC, $\beta$ & 0.1$\to$0.6\\\hline
Antenna type & Omni-directional\\\hline
Channel realization & $10^3$ \\\hline
Minimum data rate $R_{\min}$ & 0.1$\to$1.0 b/s/Hz\\\hline
Pathloss exponent & 3 \\\hline
Bandwidth & 1 Hertz \\\hline
Noise power density, $\sigma^2$ & 0.001 \\\hline
Permitted error value, $\epsilon$ & 0.001 \\\hline
Circuit power  & 5 dBm \\
\hline 
\end{tabular}
\end{table}
\section{Numerical Results and Discussion}
\begin{figure}[!t]
\hspace{0mm}
\includegraphics[width=90mm,height=60mm,trim= 100mm 0 100mm 0, clip=true]{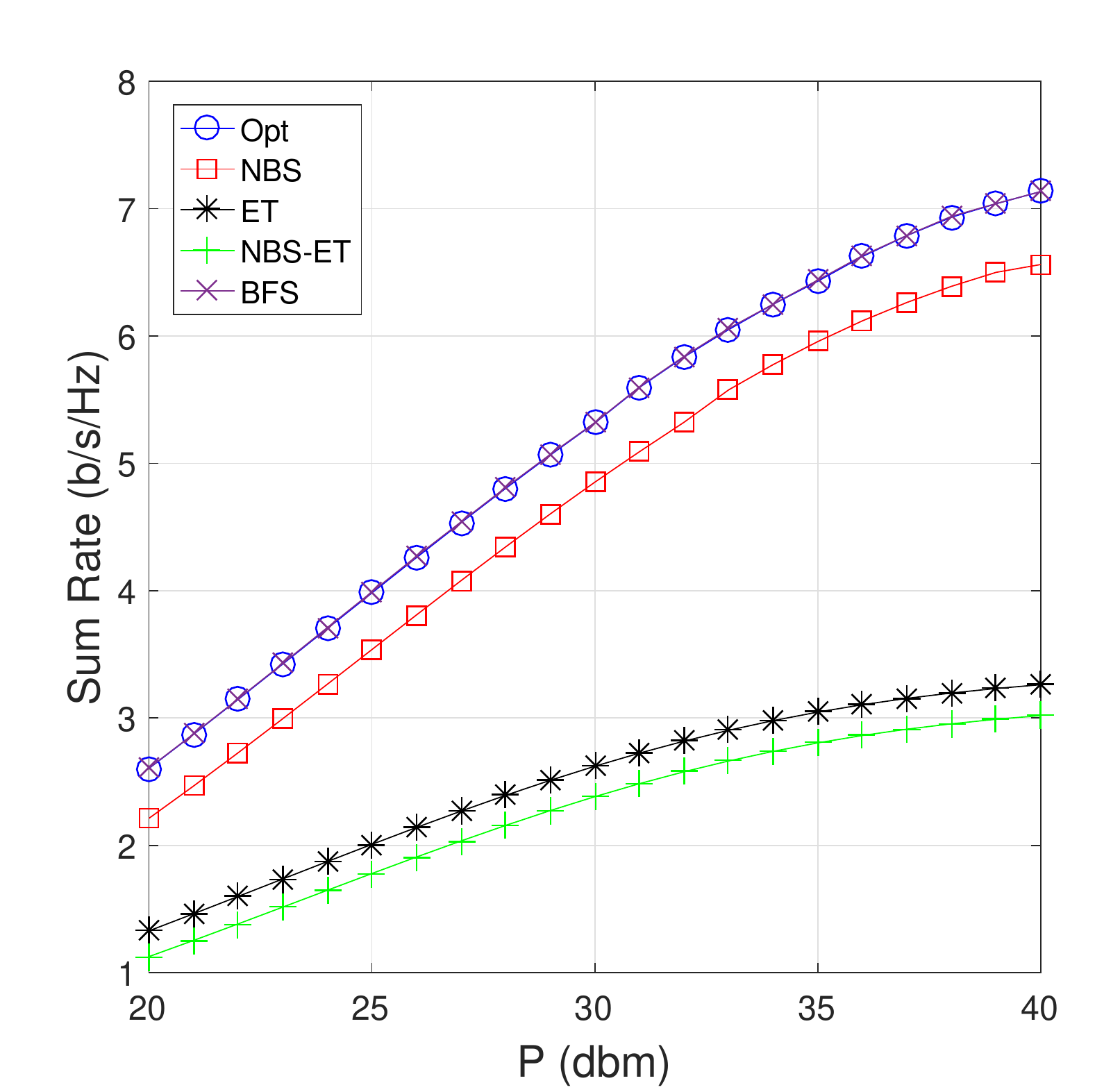}
\caption{The impact of increasing power $P$ at the BS on the Sum Rate of the system.}
\label{f3}
\end{figure}
\begin{figure}[!t]
\hspace{0mm}
\includegraphics[width=90mm,height=60mm,trim= 100mm 0 100mm 0, clip=true]{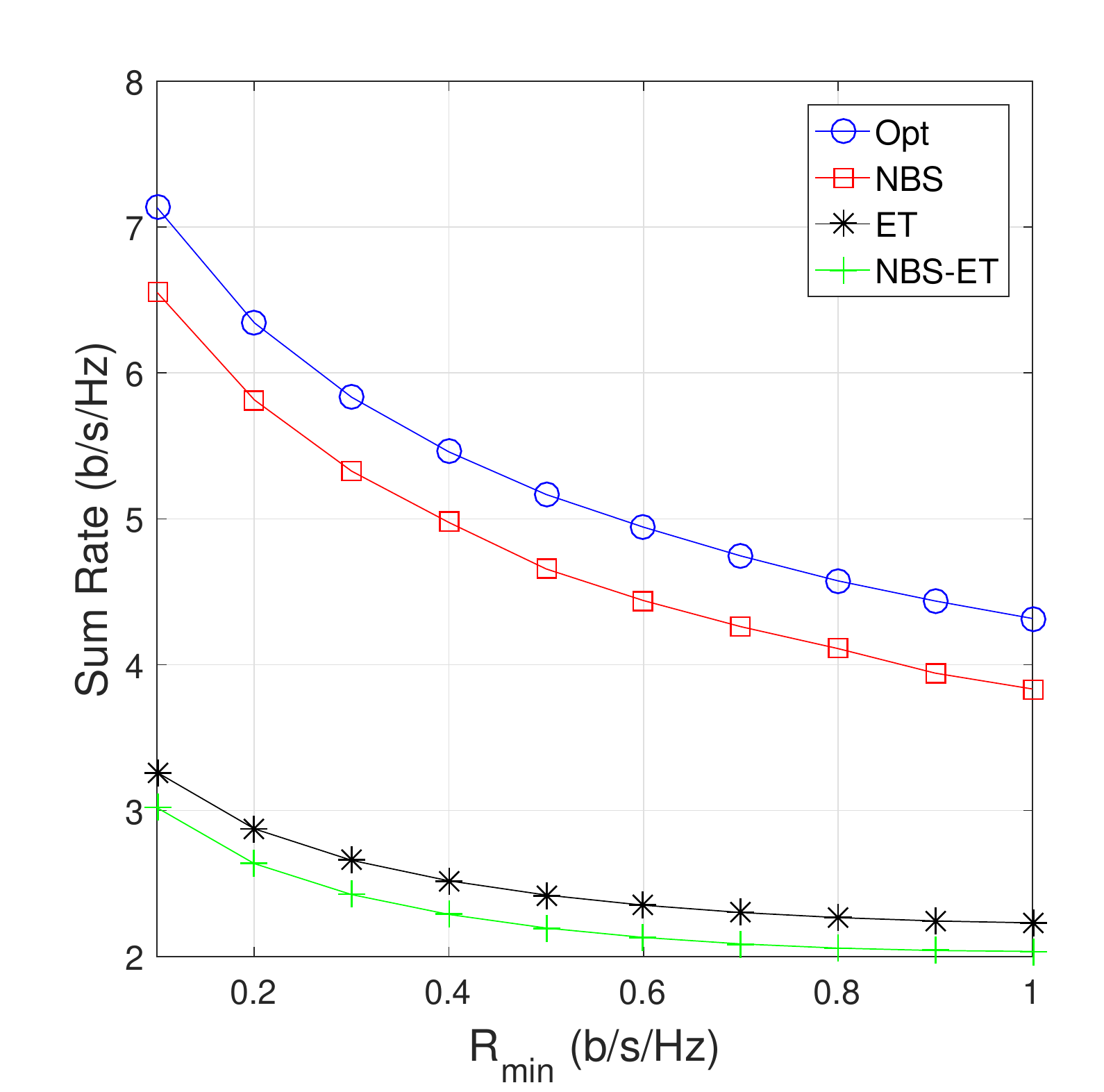}
\caption{The impact of increasing rate requirement $R_{min}$ on the Sum Rate of the system.}
\label{f2}
\end{figure} 
In this section, we present and discuss the simulation results. For the simulations, we have taken $\sigma^2=0.001$, $R_{min}=0.1$, $P=40$ dBm, $Pr_{max}=20$ dBm, $\beta=0.1$, $\epsilon=0.001$ and $\Delta=0.01$, until specified otherwise. We use Monte Carlo simulation to obtain the average results. The detailed of simulation parameters is also provided in Table II. In this work, we consider 1 Hertz bandwidth over each link. More specifically, we compute the system sum rate per Hertz. Moreover, our optimization framework is independent to the effect of bandwidth/frequency, any bandwidth can be efficiently used to obtain the simulation results. We provide the comparison of four systems \textbf{Opt}, \textbf{NBS}, \textbf{ET} and \textbf{NBS-ET}, respectively. More specifically, \textbf{Opt} refers to the proposed backscattering aided optimization framework. Then, \textbf{NBS} is the system without backscattering tag and we use the same proposed optimization technique to optimize all the system parameters. In \textbf{ET} scheme, all the parameters are optimized for the fix value of time allocation i.e. $T=0.5$. The \textbf{NBS-ET} scheme signify a system with equal time allocation with no backscattering tag in the system.

Since the considered problem has never been solved in the literature before. Thus, to evaluate the performance of the proposed frameworks we compare the performance with a brute force search \textbf{BFS} technique, where the value of the objective function is checked for each possible value of the optimization variables ($\phi_1,\phi_2,\Lambda,P_r,T$). This is a very slow technique, thus, it can not be employed in practical systems. However, this technique provides an optimal solution which can be used to evaluate the performance of the proposed optimization frameworks. The effect of increasing $P$ on the sum rate is shown in Fig. \ref{f3}. It is clear from the figure that the \textbf{Opt} provides the same results as \textbf{BFS} technique. This proves the optimality of the proposed solution technique. It can be seen that an increase in the value of $P$ results in increasing the sum rate of the system as more power becomes available for the transmission. This is because the objective function is a concave monotonically increasing function of $P$. Further, for fixed $R_{min}$ when the value of available power is increased, the difference in the rates of equal time schemes and optimal time schemes also increases. The reason behind this is that, at small value of $P$ if we reduce the time allocated for the transmission of a user, then the rate requirement might not be satisfied. However, when $P$ is increased, more power is allocated for the transmission and so the parameter $T$ becomes more flexible. Thus, optimizing $T$ gives us much better rate compared to equal $T$ cases (\textbf{ET} and \textbf{NBS-ET}).

\begin{figure}[!t]
\hspace{0mm}
\includegraphics[width=90mm,height=60mm,trim= 100mm 0 100mm 0, clip=true]{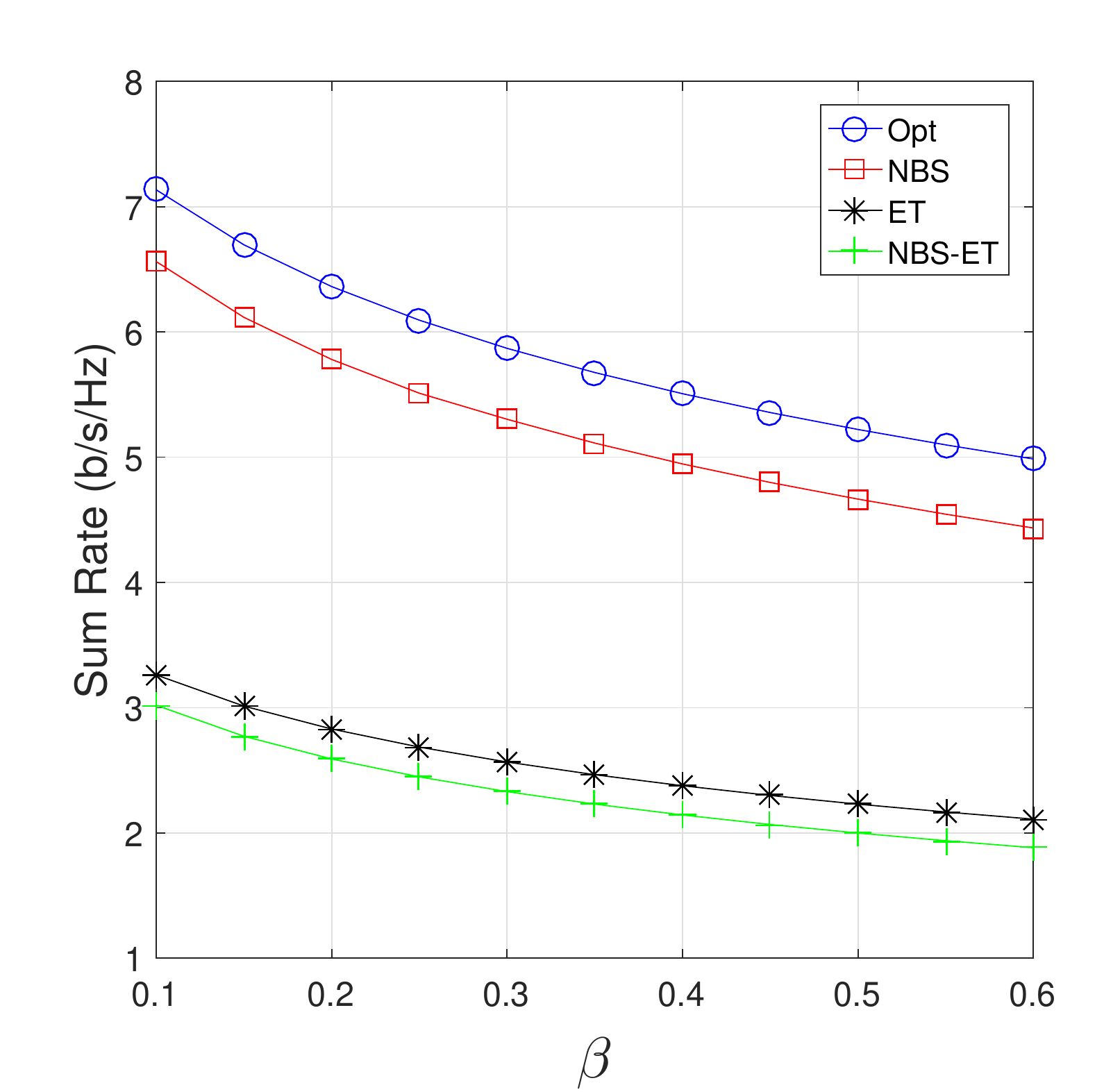}
\caption{The effect of SIC decoding errors $\beta$ on Sum Rate of the system.}
\label{f4}
\end{figure}
The impact of increasing the minimum rate requirement of each user on the sum rate of the system is shown in Fig. \ref{f2}. It can be seen that for each case, an increase in $R_{min}$ results in decreasing the overall rate of the system. This is because when $R_{min}$ is increased, more resources are required to meet the rate requirement of all users, so the optimization becomes more tightly bounded. Thus, the optimization is performed for comparatively less resources and the sum rate decreases. The figure shows that the best performance is provided by the proposed \textbf{Opt} scheme. This is because in \textbf{Opt}, all the resources are being optimized and the users benefit from the additional gain due to backscattering tag. In \textbf{NBS}, as the system has no backscattering tag, the SINR of the users is less compared to the \textbf{Opt} case. Hence, the sum rate of the system is less compared to \textbf{Opt}. Similarly, the result shows that optimizing $T$ has a significant impact on the performance, as the sum rates offered by \textbf{ET} and \textbf{NBS-ET} are far less compared to \textbf{NBS} and \textbf{Opt}. 

The Fig. \ref{f4} shows that larger value of imperfect SIC $\beta$ results in smaller sum rate of the system. With an increase in the interference faced by the $U_1$, the amount of available resources required by the user to meet the rate requirement also increase. Thus, the sum rate of the system decreases. All the schemes provide better performance when the value of $\beta$ is small. Another point worth mentioning here is that the gap in the rates provided by the backscattering system and networks with no backscattering increases if we optimize $T$. As for the same amount of transmission power the backscattering increases the SINR of the users as compared to the SINR in no backscattering case. Hence, the users in the backscattering systems can achieve the minimum required rate at comparatively smaller values of allocated power and time. Moreover, for backscattering system, the benefit of optimizing $T$ also increases. This behavior is also consistent in all the simulation results.  

\begin{figure}[!t]
\hspace{7mm}
\includegraphics[width=75mm,height=55mm,trim= 100mm 0 100mm 0, clip=true]{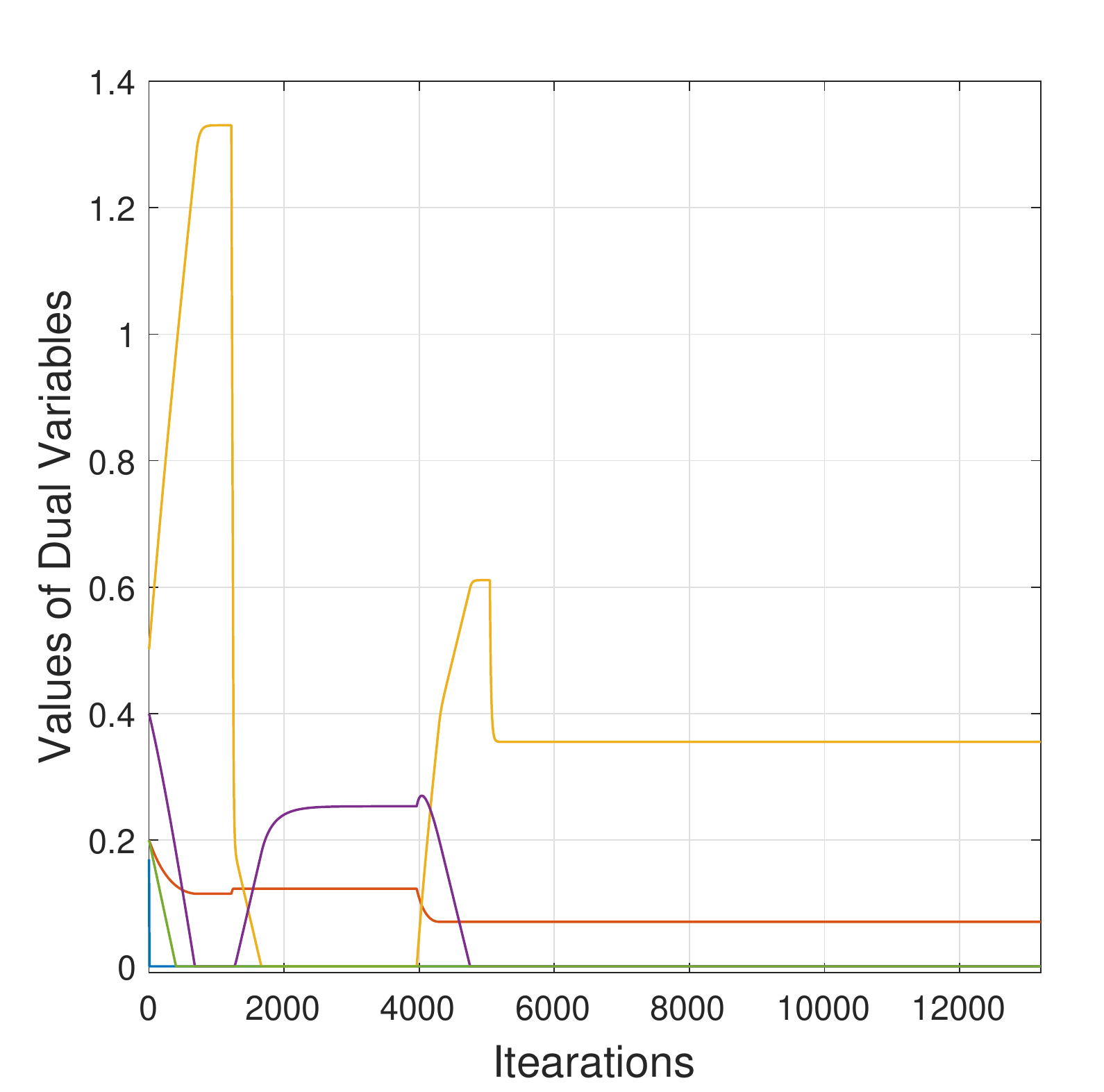}
\caption{Convergence of the dual variables involved in the proposed scheme.}
\label{f5}
\end{figure}

\begin{figure}[!t]
\hspace{3mm}
\includegraphics[width=75mm,height=55mm,trim= 90mm 0 100mm 0, clip=true]{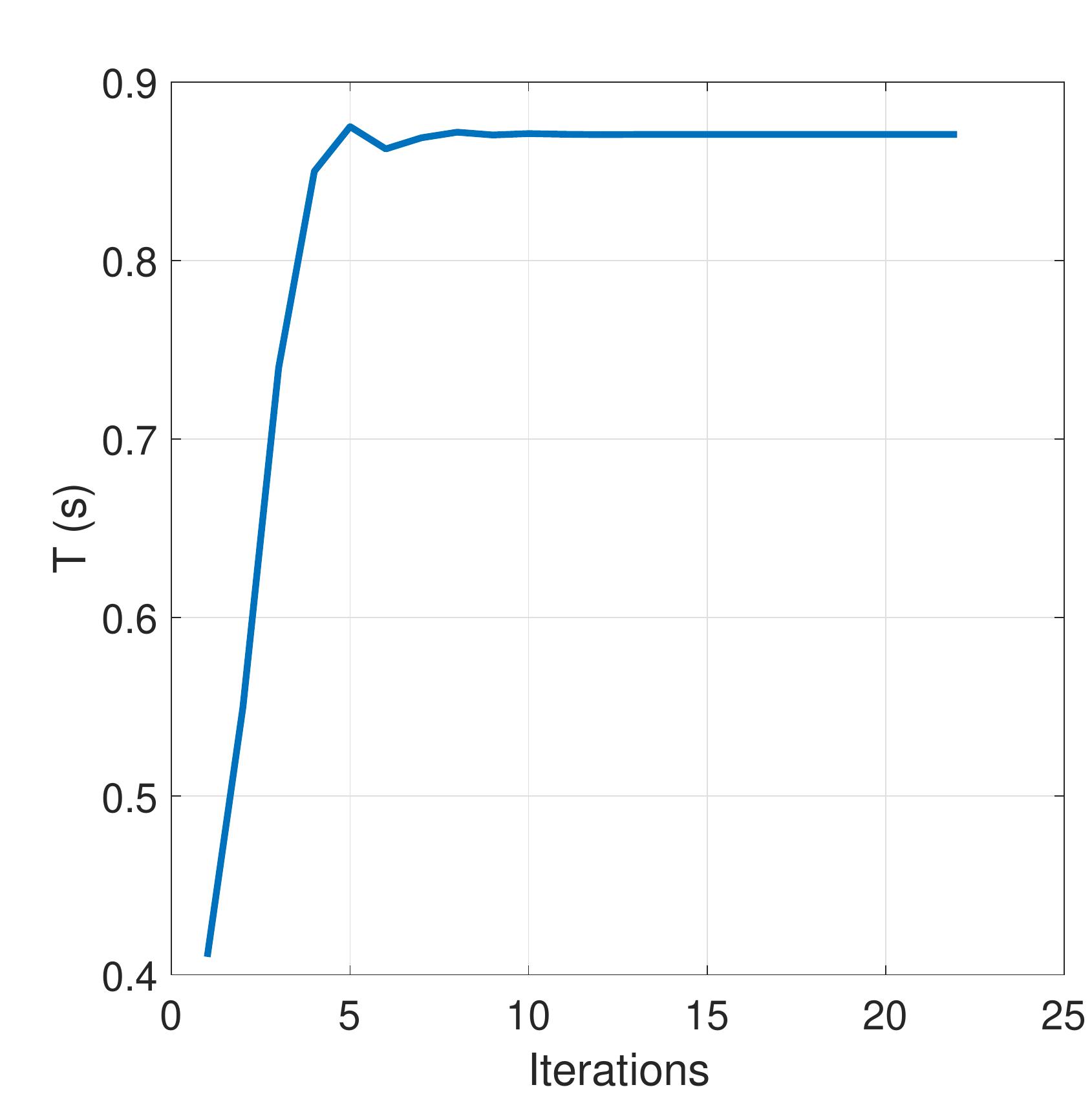}
\caption{Convergence of time involved in the proposed scheme.}
\label{f6}
\end{figure}

The convergence behavior of the proposed \textbf{Opt} framework is shown in Fig. \ref{f5}. In the \textbf{Opt} framework, when the dual variables converge, the solution is returned to the bisection method as shown in Algorithm 1. After this, the bisection method provides the updated value of $\tau$. This updated $\tau$ is again used to calculate optimal $\phi_1^*,\phi_2^*,\Lambda^*,P_r^*$, where the dual variables are again updated by using subgradient method. Due to this alternate optimization, the dual variables are updated several times till the optimal value of $\tau$ is reached by the bisection method. This behavior is clear from the Fig. \ref{f5}, e.g. at iterations $t=4000$ once the dual variables converge, the value of $\tau$ is updated by the bisection method, and the process of dual variable update starts again. However, it is clear from the figure that after a certain number of iterations, the optimal values of all the variables are reached and so the dual variables converge for iterations $t>6000$. In addition, the convergence of variable $T$ is shown in Fig. \ref{f6}. It can be seen that the bisection method provides much faster convergence compared to the dual variables in Fig. \ref{f5}.

 \begin{figure*}[!t]
\begin{align}
&\theta=f_{1} g_{3} P (g_{2} P \!+\! \sigma^2) (g_{2} \Lambda P \!+\! \sigma^2) (g_{1}^2 \Lambda^2 \nu P^2 \!+\!   g_{1} (\Lambda \Gamma \nu + \overline{\Gamma} (-\beta + (-1 \!+\! \beta) \Lambda) \mu) P \sigma^2 + (\Lambda + \Lambda \lambda_{1} + \mu - \Lambda \mu) \sigma^4) T   \nonumber\\& + (\beta g_{1} \overline{\Gamma} P - g_{1} \Lambda P - \sigma^2)(g_{1} P + \sigma^2) (g_{1} \Lambda P + \sigma^2) (f_{2} g_{3} \overline{\Gamma} (1 + \lambda_{2}\! -\! \mu) P \sigma^2 T \!+\! (g_{2} P + \sigma^2) (g_{2} \Lambda P + \sigma^2)   \zeta_{1}),\label{n13}
\end{align}\hrulefill
\begin{align}
&\theta_{1}=g_{3} P (f_{1}^2 g_{3} \Lambda P (g_{2} P + \sigma^2) (g_{2} \Lambda P \!+\! \sigma^2) (2 g_{1} \Lambda \nu P + (\Gamma+\! \lambda_{1} \!+\! \Lambda \lambda_{1} + \mu - \Lambda \mu) \sigma^2) T + f_{2} (\beta g_{1} \overline{\Gamma} P - g_{1} \Lambda P - \sigma^2) (g_{1} P + \sigma^2) \nonumber\\& (g_{1} \Lambda P\!+\! \sigma^2) (2 g_{2} \Lambda P + \sigma^2 \!+\! \Lambda \sigma^2) \zeta_{1} \!+\! f_{1} (f_{2} g_{3} P (2 g_{1}^2 g_{2} \Lambda^3 \nu P^3 + g_{1} \Lambda (2 \beta \overline{\Gamma}^2 (g_{1} (1 + \lambda_{2} - \mu) \!+\! g_{2} \mu) \!+\! \Lambda (2 g_{2} (1 \!+\! \Lambda + \lambda_{1} + \Lambda \lambda_{1} \nonumber\\& + \mu - \Lambda \mu) + g_{1} (4 + \lambda_{1} + 3 \lambda_{2} - 3 \mu + \Lambda (-2 + \lambda_{1} - 3 \lambda_{2} + 3 \mu)))) P^2 \sigma^2 + (\beta g_{1} \overline{\Gamma}^2  \Gamma (1 + \lambda_{2})\! +\! \Lambda (2 g_{2} (\Lambda \!+\! \Lambda \lambda_{1} \!+\! \mu - \Lambda \mu) \!\nonumber\\& +\! g_{1} (5 \!+\! \lambda_{1} + 4 \lambda_{2} - 3 \mu + \Lambda (\Lambda (-1 + \lambda_{1} - 2 \lambda_{2} + \mu) + 2 (\lambda_{1} - \lambda_{2} + \mu))))) P \sigma^4 \!+\! (1 \!+ \lambda_{2} + \Lambda (2 + \lambda_{1} + \lambda_{2} - \mu + \Lambda (-1 + \lambda_{1}\nonumber\\&  - 2 \lambda_{2} + \mu))) \sigma^6) T \!+\! (g_{2} P \!+\! \sigma^2) (g_{2} \Lambda\! P +\! \sigma^2) (\beta g_{1} \overline{\Gamma} P (2 g_{1} \Lambda P \!+\! \sigma^2 \!+\! \Lambda \sigma^2) \!-\! (g_{1} \Lambda P \!+\!\sigma^2) (3 g_{1} \Lambda P \!+\! \sigma^2 \!+\! 2 \Lambda \sigma^2)) \zeta_{1})),\label{n14}
\end{align}\hrulefill
\begin{align}
&\theta_{2}=g_{3}^2 P^2 (f_{1}^3 g_{3} \Lambda^2 \nu P (g_{2} P + \sigma^2) (g_{2} \Lambda P + \sigma^2) T + f_2^2 \Lambda (\beta g_{1} \overline{\Gamma} P - g_{1} \Lambda P - \sigma^2) (g_{1} P + \sigma^2) (g_{1} \Lambda P + \sigma^2) \zeta_{1} + f_{1}^2 \Lambda (f_2 g_{3} P  \nonumber\\&(\sigma^2 (2 g_{2} \Lambda(\nu  + \Lambda (\nu - \mu) + \mu) P + (3 + \lambda_{1} + 2 \lambda_{2} - \mu + \Lambda (1 + (1 + \Gamma) \lambda_{1} - \Gamma \lambda_{2} + \mu)) \sigma^2) + g_{1} P (4 g_{2} \Lambda^2 \nu P \nonumber\\& + (\beta \overline{\Gamma}^2 (1 + \lambda_{2} - \mu) + \Lambda (5 + 2 \lambda_{1} + 3 \lambda_{2} - 3 \mu + \Lambda (-1 + 2 \lambda_{1} - 3 \lambda_{2} + 3 \mu))) \sigma^2)) T + (g_{2} P + \sigma^2) (g_{2} \Lambda P + \sigma^2) \nonumber\\&(\beta g_{1} \overline{\Gamma} P - 3 g_{1} \Lambda P - (1 + \Gamma) \sigma^2) \zeta_{1}) + f_{1} f_2 (f_2 g_{3} \Lambda P (g_{1}^2 \Lambda^2 \nu P^2 + g_{1} (\Lambda \Gamma \nu + \overline{\Gamma} (-\beta + (-1 + \beta) \Lambda) \mu) P \sigma^2 + (\Lambda + \Lambda \lambda_{1}  \nonumber\\& + \mu - \Lambda \mu) \sigma^4) T + (2 g_{2} \Lambda P + \sigma^2 + \Lambda \sigma^2) (\beta g_{1} \overline{\Gamma}  P (2 g_{1} \Lambda P + \sigma^2 + \Lambda \sigma^2)- (g_{1} \Lambda P + \sigma^2) (3 g_{1} \Lambda P + \sigma^2 + 2 \Lambda \sigma^2)) \zeta_{1})),\label{n15}
\end{align}\hrulefill
\begin{align}
&\theta_{3}=f_{1} g_{3}^3 \Lambda P^3 (f_{2}^2 (\beta g_{1} \overline{\Gamma} P (2 g_{1} \Lambda P \!+\! \sigma^2 \!+\! \Lambda \sigma^2) \!-\! (g_{1} \Lambda P \!+\! \sigma^2) (3 g_{1} \Lambda P + \sigma^2 + 2 \Lambda \sigma^2)) \zeta_{1} + f_{1}^2 \Lambda (f_{2} g_{3} P (2 g_{2} \Lambda (1 + \lambda_{1}) P \!\nonumber\\& +\! (2 + \lambda_{1} + \Lambda \lambda_{1} + \lambda_{2} -  \Lambda \lambda_{2} + \overline{\Gamma} \mu) \sigma^2) T - (g_{2} P + \sigma^2) (g_{2} \Lambda P + \sigma^2) \zeta_{1}) + f_{1} f_{2} (f_{2} g_{3} \Lambda P (2 g_{1} \Lambda \nu P + (1 + \Lambda + \lambda_{1} + \Lambda \lambda_{1} \nonumber\\& + \mu - \Lambda \mu) \sigma^2) T + (2 g_{2} \Lambda P + \sigma^2 + \Lambda \sigma^2) (g_{1} (-\beta + (-3 + \beta) \Lambda) P - (2 + \Lambda)    \sigma^2) \zeta_{1})),\label{n16}
\end{align}\hrulefill
\begin{align}
\theta_{4}=f_{1}^2 f_{2} g_{3}^4 \Lambda^2 P^4 (f_{1} f_{2} g_{3} \Lambda \nu P T - f_{1} (2 g_{2} \Lambda P + \sigma^2 + \Lambda \sigma^2) \zeta_{1} + f_{2} (\beta g_{1} \overline{\Gamma} P - 3 g_{1} \Lambda P - (1 + \Gamma) \sigma^2) \zeta_{1}),\label{n17}
\end{align}\hrulefill 
\end{figure*}

\section{Conclusion}
Backscatter communication and NOMA are two promising technologies for upcoming 6G networks due to high energy and spectral efficiency.
This paper has provided the resource management framework for backscatter-aided cooperative NOMA network under imperfect SIC decoding. In particular, time allocation, power loading at BS and cooperative user, and reflection coefficient of the backscatter tag have been simultaneously optimized to maximize the sum rate of cooperative NOMA system. Closed-form solutions have been calculated by dual theory and KKT conditions. The numerical results show the efficiency of the proposed framework. Further, the results make it clear that optimizing time allocation along with power loading is very important because it significantly enhances the performance of the system, however, joint optimization of time with other optimization parameters are usually ignored in literature because of the increased complexity. 

Our proposed framework can be extended in many ways. For example, it can be extended to multi-cell NOMA cooperative communication. In that case, interference due to neighboring BSs and backscatter tags will be taken into account. This will make the problem more interesting and hard. Besides, multiple backscatter tags can also be considered in one cell to maximize spectral and energy efficiency. Further, reconfigurable intelligent surfaces is emerging 6G technology and can be used in the existing model to improve the received signal strength of the far user and replace unreliable near user cooperation. These important yet solved problems will be investigated in the future.

\section{Appendix A}
This section provides the values of $\theta,\theta_1,\theta_2, \theta_3$ and $\theta_4$ as (\ref{n13}), (\ref{n14}), (\ref{n15}) and (\ref{n16}):

where in (\ref{n13}), (\ref{n14}), (\ref{n15}) and (\ref{n16}), the values of $\Gamma=1+\Lambda$, $\overline{\Gamma}=-1+\Lambda$, and $\nu=1+\lambda_1$, respectively.

\bibliographystyle{IEEEtran}
\bibliography{zainref}

\end{document}